\documentclass[twocolumn]{aastex631}

\newcommand{\md}{\mathrm{d}}

\newcommand{\nn}{\nonumber}
\def\p{\partial}

\newcommand{\bw}{\mathbf{w}}

\newcommand{\jmin}{j_\mathrm{min}}
\newcommand{\jmax}{j_\mathrm{max}}

\newcommand{\tsec}{t_\mathrm{sec}}

\shorttitle{Phase mixing of wide binaries}
\shortauthors{Hamilton}
\graphicspath{{./}{figures/}}

\begin{document}

\title{On the phase-mixed eccentricity and inclination distributions of wide binaries in the Galaxy}

\author[0000-0002-5861-5687]{Chris Hamilton}
\affiliation{Institute for Advanced Study, Einstein Drive, Princeton, NJ 08540}





\begin{abstract}

Modern observational surveys allow us to probe the distribution function (DF) of the Keplerian orbital elements of wide binaries in the Solar neighbourhood. This DF exhibits non-trivial features, in particular a superthermal distribution of eccentricities for semimajor axes $a\gtrsim 10^3$AU.
To interpret such features we must first understand how the binary DF
is affected by dynamical perturbations, which typically fall into two classes: (i) stochastic kicks from passing stars, molecular clouds, etc. and (ii) secular torques from the Galactic tide.
Here we isolate effect (ii) and 
calculate the time-asymptotic, phase-mixed DF for an
ensemble of wide 
binaries under quadrupole-order tides. 
For binaries wide enough that the phase-mixing assumption is valid, none of our results depend explicitly on semimajor axes, masses, etc.
We show that unless the initial DF is both isotropic in binary orientation \textit{and} thermal in eccentricity,
then the final phase-mixed DF is always both anisotropic and non-thermal. However, the only way to produce a superthermal DF under phase mixing is for the initial DF to itself be superthermal.

\end{abstract}

\section{Introduction}
\label{sec:Introduction}

Measuring the Keplerian orbital elements of individual wide binaries in the Galaxy is a very difficult observational task because of the extremely long orbital periods involved.
However, in recent years the arrival of GAIA data has meant that \textit{statistical} measurements of the distribution function (DF) of binary orbital elements are now possible.  In particular, both \citet{Tokovinin2020-go} and \citet{Hwang2021-ke} have analyzed the distribution of relative position and velocity vectors of binary components, which contains statistical information about binary eccentricities, and thereby
claimed detection of a superthermal eccentricity distribution ($P(e) \propto e^\alpha$ with $\alpha > 1$) for binaries with projected separations $\gtrsim 10^3$AU in the Solar neighbourhood.
The origin of this superthermal distribution is unexplained, but is  presumably affected by the birth distribution of binaries, and the subsequent dynamical perturbations those binaries experience due to (i) scattering from passing stars, molecular clouds, and so on, and (ii) the torquing effect of Galactic tides.

How do we expect effects (i) and (ii) to drive the eccentricity DF?  For effect (i), conventionally it is thought that a sufficient number of strong scatterings will drive the binary ensemble to uniformity in orbital phase space, leading to a thermal eccentricity DF $P(e) = 2e$ (e.g. \citealt{Heggie1975-so,Binney2008-ou} --- though see
\citealt{Geller2019-ks}, who argue that the timescale for this `thermalisation' can be prohibitively long).
\citet{Stone2019-lz} found that chaotic three-body interactions produce a surviving population of binaries that is somewhat superthermal in eccentricity.
Meanwhile, a succession of weak, distant encounters causes binary eccentricity to undergo a random walk; the DF diffuses until it settles on a steady state that prefers low eccentricities, approximately $P(e)\propto e^{-0.16}$
\citep{Hamers2019-kg}.
Conversely --- and perhaps more importantly for the wide, soft binaries we have in mind here --- \citet{Collins2008-ah} found that under 
impulsive encounters binaries perform not random walks but Levy flights
in both eccentricity and inclination.
Unfortunately one cannot extract a useful steady state DF from their study as it applied to near-circular binaries only.

The impact of Galactic tides (effect (ii)) upon the eccentricity DF has not been studied in much detail. An exception is
\citet{Penarrubia2021}, who simulated the evolution of very wide binaries formed in stellar streams.
The initial DF he considered resulted from the binary formation process, and was initially almost thermal with a small deficit of highly eccentric binaries. He found that after $3$ Gyr the eccentricity DF of these binaries was even closer to thermal (with similar results when including kicks from passing substructure, i.e. combining effects (i) and (ii)).
Aside from this, the impact of Galactic tides upon wide binaries with an arbitrary initial DF is still an open question.

The time-evolution of \textit{individual} binaries under Galactic tidal perturbations is well understood as a secular phenomenon akin to the Lidov-Kozai (LK) mechanism that operates in hierarchical triples (\citealt{Lidov1962-tu,Kozai1962-ck}).
In LK theory, an inner binary can be torqued by its tertiary perturber into undergoing eccentricity and inclination oscillations on long timescales. In the doubly-averaged, test-particle, quadrupole-tide limit\footnote{See \citet{Naoz2016-qj} for details of this terminology.} the LK dynamics are governed by a simple Hamiltonian such that the binary orbital elements evolve along a 1-dimensional contour of constant Hamiltonian in phase space.
\citet{Heisler1986-eb} showed that similar secular behaviour arises when a wide binary (in their case consisting of the Sun and an Oort comet) is perturbed by the Galactic tide.  More recently, 
\citet{Hamilton2019-jn,Hamilton2019-zl}
generalised the LK and \citet{Heisler1986-eb} studies to cover \textit{any} binary in \textit{any} axisymmetric potential (see also \citealt{Brasser2006-vp,Mikkola2006-oa,Petrovich2017-py}; a further extension to triaxial potentials was studied by \citealt{Bub2020-sb}).  They derived an effective secular Hamiltonian $H_\Gamma$ which encompasses all information about the external potential and the binary's barycentric orbit around that potential in a dimensionless number $\Gamma$.  They showed that the LK Hamiltonian is recovered exactly in the limit $\Gamma=1$, while the \citet{Heisler1986-eb} problem corresponds to $\Gamma=1/3$, and they mapped out the details of the resulting dynamics for arbitrary $\Gamma$. 

Thus, there is no shortage of analytical studies of secular dynamics of individual binaries 
in external potentials.  These calculations have been used in many semi-analytical/numerical studies and population synthesis calculations of LK (and similar) evolution (e.g. \citealt{Fabrycky2007-tu,Antonini2012-fv, Stephan2016-oz,Hamilton2019-mq,Grishin2021-zb}), which were mostly concerned with using eccentricity excitation to produce exotic phenomena such as black hole mergers, hot Jupiters and blue stragglers.  However, in the age of high precision missions like GAIA it is becoming possible to measure the dynamical properties of an entire ensemble of binaries.

In this paper we develop a new tool for understanding the orbital element distribution of wide stellar binaries by calculating the time-asymptotic, `phase-mixed' DF of binaries undergoing secular dynamical evolution governed by the Hamiltonian $H_\Gamma$. The key idea is that on long timescales, the final coarse-grained distribution of binaries in phase space can be calculated by smearing the initial distribution uniformly along individual Hamiltonian contours.
 This idea
 extends back at least as far as the classic work by \citet{ONeil1965-uy}, who used it to calculate the phase-mixed velocity distribution of electrons trapped in an electrostatic plasma wave.  It has been used recently in the galactic dynamics context to calculate the DF of stars and dark matter particles that are trapped by various Galactic resonances \citep{Binney2016-zh,Monari2017-tu,Chiba2021-sg}.

The rest of this paper is organised as follows.
In \S\ref{sec:Dynamical_framework} we introduce our notation and write down the expression for the Hamiltonian $H_\Gamma$ governing the dynamics of a single binary.  In \S\ref{sec:DF} we turn to a statistical description and show how to calculate the time-asymptotic, phase-mixed distribution function (DF) of binaries in phase space for arbitrary $\Gamma$ and initial DF.
In \S\ref{sec:Results} we show the resulting final eccentricity and inclination distributions for several example cases.
We discuss our results in \S\ref{sec:Discussion} and conclude in \S\ref{sec:Conclusion}.


\section{Secular dynamics of a single binary}
\label{sec:Dynamical_framework}

Here we recap some results and notation from \citet{Hamilton2019-jn,Hamilton2019-zl} concerning the secular dynamics of one binary.

Consider a binary with component masses $m_1$, $m_2$, orbiting in a smooth, axisymmetric Galaxy potential $\Phi$ whose symmetry axis is $Z$. Let $(X,Y)$ describe the Galactic plane perpendicular to $Z$.
Then on long timescales the binary's barycentric (`outer') orbit usually fills an axisymmetric torus \citep{Binney2008-ou}. 
The binary's internal (`inner') orbital motion traces a Keplerian ellipse, described by the usual orbital elements \citep{Murray1999-zm}: semi-major axis $a$, eccentricity $e$, inclination $i$ (relative to the $(X,Y)$ plane), longitude of the ascending node $\Omega$ (relative to the $X$ axis), argument of pericentre $\omega$ and mean anomaly $\eta$.
Crucial for our purposes is the introduction of Delaunay actions $L=\sqrt{G(m_1+m_2) a}, J=L\sqrt{1-e^2}$, and $J_z = J\cos i$, and their conjugate angles $\eta$, $\omega$ and $\Omega$, as well as the dimensionless variables
\begin{eqnarray}  
\label{eqn:def_jz}
j \equiv J/L = \sqrt{1-e^2},
\\
j_z \equiv J_z/L = (1-e^2)^{1/2}\cos i.
\label{eq:AMdefs}
\end{eqnarray} 
Clearly, $j$ must obey $\vert j_z \vert \leq j\leq 1$
to be physically meaningful at a fixed $j_z$.

We assume that the outer orbit is given and fixed (i.e. there is no relaxation of outer orbits).
The evolution of the binary's inner orbit is then dictated by the mutual Newtonian gravitational attraction of the binary components and the perturbing tidal influence of the Galactic potential $\Phi$. Expanding the tides to quadrupole order and averaging over the inner and outer orbital motion we can show that the binary undergoes oscillations in $\omega, j$ at a fixed $j_z$ and $L$.  Precisely, the binary moves around the $(\omega, j)$ plane on contours of constant dimensionless Hamiltonian
\begin{eqnarray}  
\nn H_\Gamma(\omega, j, j_z) \equiv j^{-2}\left[ (j^2 - 3\Gamma j_z^2)( 5-3j^2) \right. \\ \left. \,\,\,\,\,\,\, - 15\Gamma(j^2-j_z^2)(1-j^2) \cos 2\omega \right]. 
\label{eqn:H} 
\end{eqnarray} 
Here the dimensionless quantity $\Gamma$ depends on the Galactic potential and the choice of outer orbit, and measures the time-averaged curvature of $\Phi$ as felt by the binary. Typical values of $\Gamma$ are in the range $(0,1)$. In particular, for binaries orbiting a thin disk we find $\Gamma=1/3$ \citep{Heisler1986-eb}.
It turns out that very high eccentricities are much more readily achieved if $\Gamma>1/5$ compared to $\Gamma < 1/5$ (\citealt{Hamilton2019-zl}).

The nodal angle $\Omega$ also evolves under secular dynamics; its equation of motion is $\md\Omega/\md t \propto \partial H_\Gamma/\partial j_z$.  However, since $H_\Gamma$ is independent of $\Omega$, none of the other quantities depend on $\Omega$ for their evolution, so it is effectively decoupled from the rest of the phase space and we will integrate it out in \S\ref{sec:DF}. 

The secular period --- i.e. the time it takes for the binary to perform the oscillation in the $(\omega, j)$ plane --- differs depending on the precise initial conditions but a reasonable estimate is
\begin{eqnarray} 
 t_\mathrm{sec} & \sim & \,\,\, T_Z^2/T_\mathrm{b}
\label{eqn:T_sec}\\ 
& \sim & 10^9 \, \mathrm{yr} \times \left( \frac{\rho_0}{0.2 M_\odot\,\mathrm{pc}^{-3}} \right)^{-1}
\left( \frac{m_1+m_2}{M_\odot} \right)^{1/2} \nn \\ &\times &\left( \frac{a}{10^4\mathrm{AU}} \right)^{-3/2},
\label{eqn:t_sec_numerical}
\end{eqnarray} 
where $T_Z$ is the period of vertical oscillations of the outer orbit in the Galactic potential, and $T_b = 2\pi \sqrt{a^3/[G(m_1+m_2)]}$ is the inner orbital period.  In the numerical estimate (\ref{eqn:t_sec_numerical}) we used the epicyclic approximation to write $(2\pi/T_Z)^2 \approx 4\pi G\rho_0$, where $\rho_0$ is the local dynamical density \citep{Widmark2019-wp}.
From the estimate (\ref{eqn:t_sec_numerical}) we see that a  wide binary ($a\gtrsim 10^4$AU) may complete multiple secular oscillations in the lifetime of the Galaxy.


\section{The phase-mixed distribution function} \label{sec:DF}

We do not observe the time evolution of individual wide  binary orbital elements. Instead, what we observe is a snapshot of the orbital element DF.
Ignoring scattering from passing stars, we expect that an initial distribution of binaries with a given $\Gamma$ and $j_z$ value will end up (on timescales long compared to 
$t_\mathrm{sec}$) uniformly distributed  (i.e. phase-mixed) along contours of $H_\Gamma(\omega, j, j_z)$
in the $(\omega, j)$ phase space.  
In this section we introduce the phase space DF (\S\ref{sec:phase_space_DF}) and demonstrate how one may calculate the time-asymptotic, phase-mixed DF for an arbitrary initial DF (\S\ref{sec:phase_mixed_DF}) and then for a DF that is initially isotropic in binary orientation (\S\ref{sec:Initially_isotropic}).


\subsection{Time-dependent distribution function}
\label{sec:phase_space_DF}
\begin{figure}
\centering
   \includegraphics[width=0.8\linewidth]{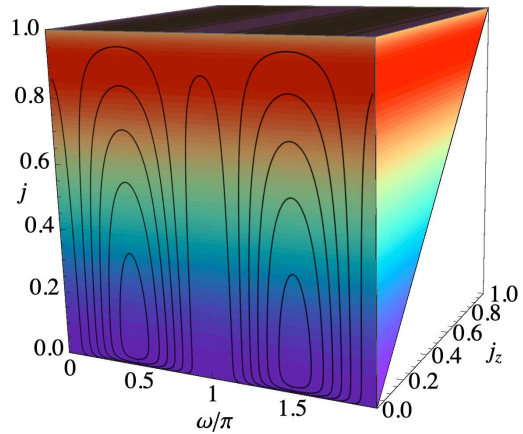}
\caption{The shape of the allowed $(\omega, j, j_z)$ phase space `wedge' at a fixed (arbitrary) value of $\Omega$, showing $j_z>0$ only.
Colors represent values of the initial DF, $\log_{10} f_0(\bw)$.
In this case $f_0$ is isotropic with Gaussian eccentricity distribution, $P_0(e) = (2\pi\sigma_e^2)^{-1/2}\exp[-(e-\mu_e)^2/2\sigma_e^2]$ where $\mu_e = 0.5$ and $\sigma_e = 0.1$.
Black contours on the front face of the wedge are of constant Hamiltonian $H_\Gamma$, in this case for $\Gamma=1/3$.
Within each constant $j_z$ `slice',
the Hamiltonian flow induced by the Galactic tide transports binaries periodically around the black contours
in the $(\omega, j)$ plane, leading to phase mixing --- see Figure \ref{fig:phase_mixed_example_1}.}
    \label{fig:phase_space_wedge}
\end{figure}

Considering only binaries whose secular periods are much shorter than their lifetime (which is not always a good assumption, see \S\ref{sec:Phase_Mixing_Discussion}), none of the results we derive will depend explicitly on
$a$, $m_1, m_2$, Galaxy mass, etc.
Instead the only variables of concern are $\Gamma$ and $\bw$ where
\begin{eqnarray}
    \bw \equiv (\omega, \Omega, j, j_z).
\end{eqnarray}
Let us therefore consider such an ensemble of binaries
all with the same value of $\Gamma$ (e.g. all on similar outer orbits in the same Galactic potential $\Phi$).
To describe this ensemble we introduce the smooth 4D probability distribution function $f(\bw,t)$, such that 
$f(\bw, t)\md \bw$ is the fraction of binaries in the phase space volume element $(\bw, \bw + \md \bw)$ at time $t$.
This DF is normalised so that 
\begin{eqnarray}
    \int_0^{1}\md j  \int_{-j}^{j}\md j_z  \int_0^{2\pi}\md \Omega \int_0^{2\pi}\md \omega \, f(\bw, t) = 1.
    \label{eqn:4D_DF_normalisation}
\end{eqnarray}
The limits on the $j_z$ integration reflect the requirement $\vert j_z \vert \leq j \leq 1$.  The shape of the 3D phase space $(\omega, j, j_z)$ at fixed (arbitrary) $\Omega$ is illustrated in Figure \ref{fig:phase_space_wedge} for $j_z>0$.

For later use we also define the 1D distribution of dimensionless angular momenta $F(j,t)$:
\begin{eqnarray}
    F(j,t) \equiv \int_{-j}^{j}\md j_z  \int_0^{2\pi}\md \Omega \int_0^{2\pi}\md \omega \, f(\bw, t),
    \label{eqn:1D_angular_momentum_DF}
\end{eqnarray}
which satisfies $\int_0^1 \md j \, F(j,t) = 1$.
Ultimately we 
care about the 1D distribution of eccentricities which we call $P(e,t)$; we can convert between $F$ and $P$ using $\vert F(j,t) \md j \vert = \vert P(e,t)\md e \vert$, i.e.
\begin{eqnarray}
   P(e,t) = \frac{e}{\sqrt{1-e^2}}F(\sqrt{1-e^2},t),
   \label{eqn:1D_DF_eccentricity}
\end{eqnarray}
One can check that $\int_0^1 \md e \,P(e,t)=1$.

Since the Galactic plane picks out a special direction it is natural to ask whether a non-trivial phase-mixed inclination distribution can arise. To calculate this we introduce the 1D DF of $\cos i$ values
%
\begin{eqnarray}
& & N(\cos i, t)
\nn\\
& & \equiv
    \int_{0}^{1} \md j  \int_0^{2\pi} \md \Omega \int_0^{2\pi} \md \omega 
    \,j f(\omega, \Omega, j, j\cos i, t),
    \label{eqn:1D_DF_inclination}
\end{eqnarray}
which satisfies $\int_{-1}^1 \md \cos i \, N(\cos i, t) = 1$.


\subsection{Phase-mixed distribution function}
\label{sec:phase_mixed_DF}
\begin{figure*}
\centering
   \includegraphics[width=0.7\linewidth,trim={0cm 0 0 0.9cm},clip]{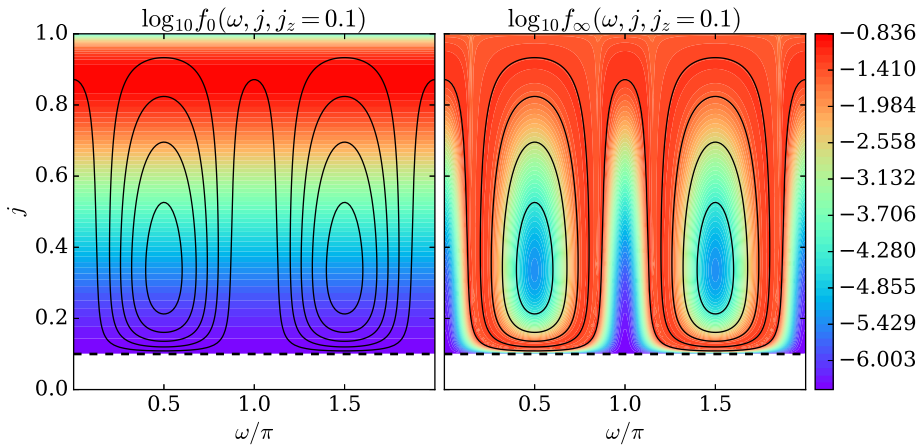}
\caption{Illustration of phase mixing. Colors show $\log_{10}f_0$ (left) and $\log_{10}f_\infty$ (right) in the $(\omega, j)$ phase space at fixed $\Gamma =1/3$ and $j_z=0.1$, for the same initial DF used in Figure \ref{fig:phase_space_wedge}.
Solid black lines show contours of constant Hamiltonian $H_\Gamma$. The black dashed line shows the lowest possible angular momentum $j=\vert j_z\vert$.}
    \label{fig:phase_mixed_example_1}
\end{figure*}

Now we wish to calculate the time-asymptotic, phase-mixed DF $f(\bw, t\to \infty) \equiv f_\infty(\bw)$ for a given $\Gamma$ and initial DF $f(\bw,t=0) \equiv f_0(\bw)$.
\footnote{Though we refer to $f_0$ as the `initial' DF, given that
 the results we derive are time-asymptotic there is nothing particularly special about $t=0$.  In other words,
 it is not important whether the binaries were all born in a single burst at $t=0$ or gradually over billions of years. What matters
is that the ensemble under consideration is sufficiently old for the phase-mixing assumption to be valid --- see \S\ref{sec:Phase_Mixing_Discussion}.} To do this, we note that individual binaries are advected around $H_\Gamma$ contours periodically by the Galactic tide.  In the canonical Delaunay phase space coordinates we are using here, Liouville's theorem tells us that these advected binaries carry with them the local phase space density $f$.
Binaries on adjacent contours have slightly different secular periods $t_\mathrm{sec}$, so $f$ is continually sheared out until its coarse-grained value reaches a steady `phase-mixed' state in which it is spread \textit{uniformly} over each contour \citep{ONeil1965-uy,Lynden-Bell1967-er,Tremaine1999-im}.

The phase-mixed DF $f_\infty(\bw)$ may therefore be calculated as follows.
First, we use $f_0$ to calculate the fraction $\mathcal{A}(\bw)$ of binaries 
that are born on the Hamiltonian phase space contour defined by $\bw$:
\begin{eqnarray}
    \mathcal{A}(\bw) \equiv \oint_{\mathcal{C}_\Gamma(\bw)} \md \lambda \, f_0(\omega'(\lambda), \Omega, j'(\lambda), j_z).
    \label{eqn:contour_population}
\end{eqnarray}
Here we have labelled this contour \begin{eqnarray}
    \mathcal{C}_\Gamma(\bw) \equiv \{ \omega',j' \, \vert \, H_\Gamma(\omega', j', j_z) = H_\Gamma(\omega, j, j_z) \},
    \label{eqn:contour_def}
    \end{eqnarray}
    and parameterised it by $\lambda$.
Next, we calculate the length $\mathcal{L}(\bw)$ of the contour $\mathcal{C}_\Gamma(\bw)$ in phase space:
\begin{eqnarray}
    \mathcal{L}(\bw) \equiv \oint_{\mathcal{C}_\Gamma(\bw)} \md \lambda.
    \label{eqn:contour_length}
\end{eqnarray}
The initial density will ultimately be smeared evenly over the full length of the contour, so the value of the phase-mixed distribution function at the location $\bw$ is simply
\begin{eqnarray}
    f_\infty(\bw)
    &= {\mathcal{A}(\bw)}/{\mathcal{L}(\bw)}.
    \label{eqn:phase_mixed_ratio}
\end{eqnarray}
It is straightforward to show that the DF constructed in this way is properly normalised, i.e. $\int \md \bw f_\infty = 1$ (equation (\ref{eqn:4D_DF_normalisation})). 
In the Appendix we go into more detail about how $f_\infty$ is calculated in practice.

As an illustration, in Figure \ref{fig:phase_mixed_example_1}
we consider binaries with $\Gamma=1/3$ and at a fixed $j_z = 0.1$, for an initially Gaussian eccentricity distribution, $P_0(e) = (2\pi\sigma_e^2)^{-1/2}\exp[-(e-\mu_e)^2/2\sigma_e^2]$ with mean $\mu_e = 0.5$ and standard deviation $\sigma_e = 0.1$. In the left panel the colored contours map the initial DF $\log_{10} f_0(\bw)$ in $(\omega, j)$ space, while the solid black contours denote lines of constant Hamiltonian $H_\Gamma$ (in fact this panel is nothing more than the $j_z=0.1$ `slice' of the 3D wedge shown in Figure \ref{fig:phase_space_wedge}).
In the right panel of Figure \ref{fig:phase_mixed_example_1} we show the resulting phase-mixed DF $f_\infty$. We see that it overlays the Hamiltonian contours precisely, and that binaries are able to spread over a large range of eccentricities.



Once we have calculated the final 4D phase-mixed DF $f_\infty(\bw)$, we can easily get the final 1D angular momentum distribution $F_\infty(j)$ by plugging $f(\bw, t) = f_\infty(\bw)$ into equation 
 (\ref{eqn:1D_angular_momentum_DF}).
The final 1D eccentricity distribution then follows from equation (\ref{eqn:1D_DF_eccentricity}) as $P_\infty(e) = e F_\infty(\sqrt{1-e^2})/\sqrt{1-e^2}$.
Similarly for inclination, the 1D phase-mixed DF $N_\infty(\cos i)$ is found by substituting $f(\bw,t) = f_\infty(\bw)$ in equation (\ref{eqn:1D_DF_inclination}). 


\subsection{Initially isotropic distributions}
\label{sec:Initially_isotropic}

A key simplification can be made if we assume that the birth DF $f_0$ is isotropic in binary orientation,  i.e. uniform in $\omega, \Omega$ and $\cos i$ ($=j_z/j$). Then $f_0$ only depends on $j$, and consequently we can relate it to the initial 1D distributions of angular momentum $F(j,t=0) \equiv F_0(j)$ and/or eccentricity 
$P(e,t=0) \equiv P_0(e)$ as follows:
\begin{eqnarray}
 f_0(\bw) = \frac{F_0(j)}{8\pi^2 j} = \frac{P_0(\sqrt{1-j^2})}{8\pi^2 \sqrt{1-j^2}}.
 \label{eqn:convert_isotropic_4D_to_1D}
\end{eqnarray}
With this our final results for the phase-mixed DF $f_\infty$ will depend only on $\Gamma$ and the choice of initial eccentricity distribution $P_0$. For the remainder of this paper we will assume $f_0$ has the isotropic property.  

One important special case to check is that of an initially completely uniform phase space distribution over all $\bw$, namely $f_0(\bw) = 1/(2\pi)^2$.
This corresponds to a thermal eccentricity distribution, $P_0(e) = P_\mathrm{thermal} \equiv 2e$, and an isotropic inclination distribution, $N_0(\cos i) = N_\mathrm{isotropic} \equiv 1/2$.  In this case $f_0$ can be pulled out of the integral in (\ref{eqn:contour_population}) and so we find from equation (\ref{eqn:phase_mixed_ratio}) that $f_\infty = f_0 = 1/(2\pi)^2$, i.e. the final phase-mixed DF is uniform also.
It follows that a population of binaries that is initially isotropic with a thermal eccentricity distribution remains so, despite the Galactic tide continually advecting individual binaries around the $(\omega, j)$ plane.  

\section{Numerical results}
\label{sec:Results}

In this section we provide results on the 1D phase-mixed eccentricity and inclination distributions, $P_\infty(e)$ and $N_\infty(\cos i)$, for different initial eccentricity distributions $P_0(e)$ (the 
initial orientations are assumed to be isotropic so $N_0(\cos i) = 1/2$ --- see \S\ref{sec:Initially_isotropic}).
We calculated these DFs numerically using the method described in the Appendix. 
We performed calculations for several different values of $\Gamma$ and found that while the results for $\Gamma > 1/5$ and $\Gamma < 1/5$ differ greatly (as expected, see \citealt{Hamilton2019-zl}), if we stick to $\Gamma>1/5$ then the results depend on $\Gamma$ only very weakly. 

For binaries whose outer orbit is confined to the midplane of a thin Galactic disk, $\Gamma\approx 1/3 > 1/5$. In fact, all binaries in the Solar neighbourhood will have $\Gamma$ not too far from $1/3$. In the rest of this work we display results exclusively for $\Gamma=1/3$, but the qualitative conclusions should hold for any sensible population of outer orbits.



\begin{figure*}
\centering
   \includegraphics[width=0.75\linewidth,trim={0cm 0.5cm 0 0.7cm},clip]{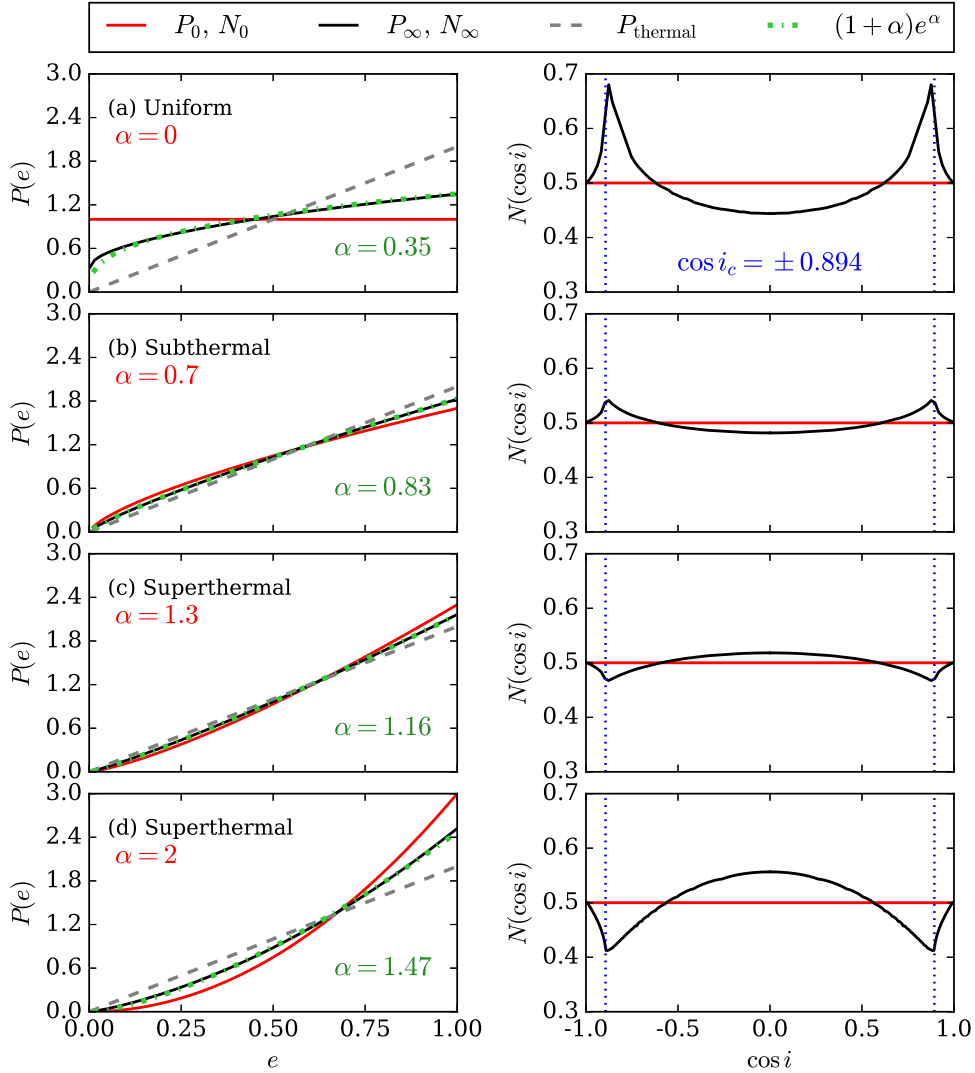}
\caption{Numerically computed 1D phase-mixed eccentricity DF $P_\infty(e)$ and inclination DF $N_\infty(\cos i)$ are shown in black
for different initial DFs shown in red.
The $P_\infty$ curves in the left panels are well-fit by power laws $P = (1+\alpha)e^\alpha$, shown with green dot-dashed lines; for comparison
we also show the thermal eccentricity distribution $P_\mathrm{thermal}=2e$
with a dashed grey line.
In the right panels we show a special value of inclination, $\cos i = \pm \vert \cos i_c \vert = \pm 0.894$, with vertical blue dotted lines --- see \S\ref{sec:Results} for details.}
    \label{fig:Eccentricity_inclination_results}
\end{figure*}

In Figure \ref{fig:Eccentricity_inclination_results} we fix $\Gamma=1/3$ and consider four different choices of initial power-law DF, $P_0(e) = (1+\alpha) e^\alpha$ with $\alpha =0,\, 0.7,\, 1.3,\, 2$ respectively:
\begin{eqnarray}
    &\mathrm{(a)}& \,\, P_0(e) = 1, \,\,\,\, 
    \mathrm{``Uniform"},
    \label{eqn:Uniform_eccentricity_DF}
    \\
    &\mathrm{(b)}& \,\, P_0(e) \propto e^{0.7}, \,\,\,\, \mathrm{``Subthermal"},
    \label{eqn:Subthermal_eccentricity_DF}
     \\
    &\mathrm{(c)}& \,\, P_0(e) \propto e^{1.3}, \,\,\,\, 
    \mathrm{``Superthermal"},
    \label{eqn:Superthermal_eccentricity_DF}
    \\
    &\mathrm{(d)}& \,\, P_0(e) \propto e^{2}, \,\,\,\, \mathrm{``Superthermal"}.
    \label{eqn:Hyperthermal_eccentricity_DF}
\end{eqnarray}
In the left panels of Figure \ref{fig:Eccentricity_inclination_results} we plot the resulting phase-mixed eccentricity distribution $P_\infty(e)$ in black, and the initial DF of choice $P_0(e)$ in red. For reference we show the thermal DF $P_\mathrm{thermal}=2e$ with a 
dashed grey line.  We additionally plot power-law fits to the black curves with green dot-dashed lines, with the best fit $\alpha$ indicated in the panel.
In the right panels of Figure \ref{fig:Eccentricity_inclination_results} we plot the corresponding phase-mixed DF of inclination $N_\infty(\cos i)$ in black, and the initial $N_0(\cos i)=1/2$ in red.
We also plot vertical blue dotted lines at $\vert \cos i \vert = \vert \cos i_\mathrm{c} \vert \equiv \sqrt{(1+5\Gamma)/10\Gamma} \approx 0.894 $, which corresponds to $\vert i \vert = \vert i_\mathrm{c} \vert \approx 26.5^\circ$.
This is the critical inclination angle below which there are no fixed points in the $(\omega, j)$ phase space for initially near-circular binaries\footnote{This is just the Galactic-tidal analogue of the classic LK result $i_c = 39.2^\circ$ (e.g. \citealt{Fabrycky2007-tu}).} --- see \S9.1 of \citet{Hamilton2019-zl}.

From these panels (and several corroborative examples not shown here) we can draw the following conclusions:
\begin{itemize}
    \item Only initially superthermal DFs remain superthermal as $t\to\infty$; the result is another superthermal DF with a slightly reduced power law index.
    
    \item Initially subthermal DFs also retain power law form and their index is increased slightly, but never beyond $1$, i.e. they remain subthermal.
    
    \item Unless a DF is initially both thermal and isotropic, it will be neither thermal nor isotropic in the $t\to\infty$ limit.
\end{itemize}

The last bullet point is worth discussing further. It 
implies that Galactic tides produce a phase-mixed DF in which eccentricities and inclinations are correlated, even if the binaries are initially distributed isotropically for any eccentricity.  
The further the initial DF is from thermal, the stronger the resulting anisotropy will be;
in examples (b)-(c) above it reaches the level of several percent,
while in examples (a) and (d) it can be tens of percent.
It is also easy to predict the angle at which the anisotropy will be most pronounced.  Roughly speaking, for Galactic tides to drive large-scale eccentricity and inclination oscillations there must be a fixed point in the $(\omega, j)$ phase space around which trajectories can librate (Figure \ref{fig:phase_mixed_example_1}).
For an initially low-$e$ binary with $\vert \cos i \vert \gtrsim \vert \cos i_\mathrm{c} \vert$ (i.e. $\vert i \vert \lesssim \vert i_\mathrm{c}\vert $), such fixed points do not exist, so these low-$e$ binaries are effectively `trapped' at low $i$.
In examples (a) and (b), there is an initial surplus of low-$e$ binaries compared to a thermal distribution; these binaries pile up at low inclinations, resulting in the maximum of $N_\infty$ around $\vert \cos i \vert = \vert \cos i_c \vert$. 
Conversely, in examples (c) and (d)  there is a deficit of initially low-$e$ binaries compared to the thermal DF, so this maximum becomes a minimum.



\section{Discussion}
\label{sec:Discussion}

\subsection{Phase mixing}
\label{sec:Phase_Mixing_Discussion}

We have assumed throughout this paper that the only perturbation binaries feel is that due to the smooth Galactic disk potential, and ignored any stochastic effects i.e. scattering from passing stars, molecular clouds, dark matter substructure, and so on.
Given that the widest binaries will certainly undergo many scattering events during a Hubble time --- and can even be disrupted by scattering (\citealt{Weinberg1987-qk, Jiang2010-fr,Penarrubia2021}) --- stochasticity cannot be ignored in a proper theory.
Nevertheless, our aim here has been to isolate the Galactic tidal effect and calculate the DF to which it drives binaries.
Its impact is to take the initial phase space DF and smear it uniformly along Hamiltonian contours (phase mixing).
A DF which is initially isotropic in orientation and thermal in eccentricity is already phase mixed, 
since it has the same value everywhere in phase space.  All non-thermal distributions must undergo some time evolution before reaching their fully phase-mixed state $f_\infty$. 


How valid is the phase mixing assumption? --- in other words, ignoring scattering, how long must one typically wait for $f_\infty$ to approximate the true DF? 
To get a rough idea we can consider two binaries with initial phase space locations $\bw_0 - \delta \bw$ and 
$\bw_0 + \delta \bw$. Expanding their equations of motion for small $\vert \delta \bw\vert \ll \vert \bw_0 \vert$
we can show that these neighbouring trajectories diverge in the $(\omega,j)$ plane on a characteristic timescale $\sim \tsec(\bw_0)$.   
Thus we can roughly state that a population of binaries must be significantly older than its typical secular period for its present-day DF to be approximately phase-mixed.
This idea is confirmed by numerical integration of the kinetic equation governing $f(\bw, t)$, which shows that the phase-mixed DF $f_\infty(\bw)$ is well-established after $\sim 5 \tsec$, with $\tsec$ given in equation (\ref{eqn:t_sec_numerical}), if we consider phase space locations $\bw$ that are not extremely close to the separatrix between librating and circulating phase space families\footnote{L. Arzamasskiy, private communication.}.
Near the separatrix this rough criterion breaks down because $\tsec$ is formally infinite there. However, this caveat applies to such a small fraction of binaries that it does not impact our results significantly.

A more complete understanding of the phase mixing process will involve following the detailed time-evolution of the 4D DF for many ensembles of binaries with different initial DFs, semimajor axes, $\Gamma$ values, etc.  It will also require dropping the secular approximation, to take account of fluctuations in the potential felt by the binary on the timescale $\sim T_Z$ \citep{Grishin2018-da, Hamilton2021-ue}.
We leave this to future work.

\subsection{Implications for wide binaries in the Galaxy}

The above discussion suggests
that binaries with $t_\mathrm{sec}$ much smaller than the age of the Galaxy --- say $a \gtrsim 10^4$AU, see equation (\ref{eqn:t_sec_numerical}) --- will be well phase-mixed. Those binaries which have $\tsec$ comparable to the Galaxy's age, say $a \sim 10^3$AU, will have undergone some phase mixing but the process is unlikely to be complete, 
and so our results cannot be naively applied to them. Instead, for these binaries one must integrate forward the kinetic equation for $f(\bw, t)$ numerically; doing so suggests that their eccentricity DF today should lie somewhere in-between the $P_0(e)$ and $P_\infty(e)$ results quoted in \S\ref{sec:Results}. Finally, for the binaries with a secular timescale much longer than the age of the Galaxy ($a < 10^3$AU) the effect of Galactic tides is negligible.
 
Observationally, various pieces of evidence regarding metallicities \citep{El-Badry2018-kw, Hwang2020-lw}, mass ratios \citep{Moe2017-iw} and eccentricities \citep{Tokovinin2020-go,Hwang2021-ke}
suggest that binaries with $a \lesssim 10^2$AU and $a\gtrsim 10^3$AU follow separate formation channels.
Let us take this literally and suppose that all binaries with $a\gtrsim 10^3$AU were formed from some channel that produced an initially superthermal DF. Then we expect that Galactic tides will not alter much the DF of $a \sim 10^3$AU binaries, whereas for $a \gtrsim 10^4$AU the DF will be close to phase mixed, i.e. still superthermal but with a slightly reduced power-law index (Figure \ref{fig:Eccentricity_inclination_results}c).
Interestingly, this is just what is observed by \citet{Hwang2021-ke} (see their Figure 6).  
Of course there are many subtleties to be addressed before one can claim this comparison between theory and observation to be precise.  To name just one, 
\citet{Hwang2021-ke} inferred their eccentricity DFs assuming an isotropic DF of binary orientations, whereas we have shown that the `special direction' picked out by the Galactic plane actually creates a non-isotropic DF in which $e$ and $i$ are correlated (and where the special values of $i$ are easily predicted).
 In principle one could
 measure the joint $e$-$i$ distribution of wide binaries 
 and use it to distinguish the impact of Galactic tides compared to other dynamical effects/formation channels. On the other hand, there will be complicated degeneracies of this distribution with that arising from chaotic evolution of very wide triple stars \citep{Grishin2021-zb}.

\section{Conclusion}
\label{sec:Conclusion}

In this paper we calculated the time-asymptotic distribution function (DF) for wide binaries under the  tidal influence of the Galactic disk. The central assumption we made was that the
secular oscillations of binary orbital elements induced by Galactic tides were sufficiently rapid for the whole population of binaries to be approximately phase-mixed. The resulting phase-mixed DFs of binary eccentricity and inclination 
are independent of the binary constituent masses, semimajor axes, the mass of the Galaxy, etc.

The two key conclusions of this work are:
(1)
Galactic tides can preserve, but not create, a superthermal eccentricity distribution.
(2)
Unless the initial DF is isotropic in angle and thermal in eccentricity, then the final phase-mixed DF is neither isotropic nor thermal.

These results may go some way to understanding the observed non-thermal (including superthermal) eccentricity distributions of wide binaries in the Solar neighbourhood \citep{Tokovinin2020-go,Hwang2021-ke}.
However, before strong conclusions can be drawn, both time-dependence and scattering from passing stars must be incorporated into the model.

\begin{acknowledgements} 

This project arose out of conversations with Hsiang-Chih Hwang and Nadia Zakamska, and I am very grateful to them both for their detailed comments on the manuscript.
I also thank Scott Tremaine and Roman Rafikov for helpful discussions on phase space mixing, Kathryn Johnston and Evgeni Grishin for comments on an earlier draft, and the anonymous referee for a careful reading.
This work was supported by a grant from the Simons Foundation (816048, CH).
\end{acknowledgements}


\appendix 
\section{Calculating the phase-mixed distribution function in practice}

As we have seen in equation (\ref{eqn:phase_mixed_ratio}), 
the phase-mixed DF $f_\infty$ is given by the ratio $\mathcal{A}(\bw)/\mathcal{L}(\bw)$, where $\mathcal{A}$ is the intial population on the Hamiltonian contour defined by $\bw$, and $\mathcal{L}$ is the length of that contour (equations (\ref{eqn:contour_population})-(\ref{eqn:contour_length})).
Both of those expressions involve integration over some abstract quantity $\lambda$ that parameterizes the contour in the $(\omega', j')$ plane. In practice we need to have some explicit
way to compute these integrals.
This is easy if we let $\lambda = t$, integrate from the time the binary is at $j'=\jmin$ to $j'=\jmax$, and multiply by $2$.
Then $\mathcal{L}$ is just the secular period of binaries on that contour, and $\mathcal{A}$ is the amount of time that any binary moving on that contour spends near $\bw$ per secular period.

Moreover, we do not need to worry about getting a precise form of $\omega'(t)$ 
if we take the initial 4D distribution $f_0$ to be isotropic in angle, i.e. independent of $\omega$.
In that case we can change the integration variable from $t \to j'$ and show that (c.f. equations (30)-(34) of \citet{Hamilton2019-zl}):
\begin{eqnarray}
    f_\infty(\bw) = \frac{\sqrt{\Delta}}{8\pi^2}
    \left[K\left(\sqrt{\frac{\jmax^2-\jmin^2}{\Delta}}\right)\right]^{-1}
    \int_{j_\mathrm{min}}^{j_\mathrm{max}} \md j' \frac{F_0(j')}{\sqrt{\vert (j_0^2-j'^2)(j_+^2-j'^2)(j'^2-j_-^2) \vert}}.
    \label{eqn:length_calculation}
\end{eqnarray}
Here $K(...)$ is an elliptical integral of the first kind, and the quantities $\jmin$, $\jmax$, $j_\pm$, $j_0$, $\Delta$ 
are all functions of $(\Gamma, \bw)$. All details of how to compute these quantities can be found in \citet{Hamilton2019-zl}.

For a given $\Gamma$ and $F_0(j)$, we compute the phase-mixed DF $f_\infty$ on a grid in the 3D $(\omega, j, j_z)$ phase space numerically using equation (\ref{eqn:length_calculation}).
Symmetry considerations mean that one can restrict the numerical calculation to $\omega \in (0,\pi/2)$ and $j_z >0$.
As a check of the code, we made sure that the output of a thermal eccentricity distribution ($P_0=2e$) is another thermal distribution to very high accuracy.  
With the $f_\infty(\bw)$ grid established we compute $P_\infty(e)$ and $N_\infty(\cos i)$ via equations (\ref{eqn:1D_angular_momentum_DF})-(\ref{eqn:1D_DF_inclination}) using a Simpson's rule integrator.

We note that we have made no reference to the angle $\Omega$, despite the fact that $\Omega$, just like $\omega$, evolves under secular dynamics.
The reason is that, since $\Omega$ is decoupled from the other variables, if initial $\Omega$ values are randomly distributed in $(0,2\pi)$ then the final DF will be uniform in $\Omega$ at a given $\omega, j, j_z$.
Thus for an initially isotropic DF, integration of $f(\bw,t)$ over $\Omega$ will always return $2\pi f(\bw,t)$.


\bibliography{bibliography}{}
\bibliographystyle{aasjournal}



\end{document}